\documentclass[aps,twocolumn,showpacs,floatfix]{revtex4}
\usepackage{amsmath}
\usepackage{amsfonts}
\usepackage{amssymb}
\usepackage{graphics}
\begin{document}
\title{Nonclassicality and decoherence of photon-subtracted squeezed states}
\author{Asoka Biswas$^{1}$ and G. S. Agarwal$^2$}
\affiliation{$^{1}$Department of Chemistry, University of Southern
California, Los Angeles, California 90089, USA\\
$^2$Department of Physics, Oklahoma State University, Stillwater,
Oklahoma 74078, USA}
\date{\today}
\begin{abstract}
We discuss nonclassical properties of single-photon subtracted
squeezed vacuum states in terms of the sub-Poissonian statistics and
the negativity of the Wigner function. We derive a compact
expression for the Wigner function from which we find the region of
phase space where Wigner function is negative. We find an upper
bound on the squeezing parameter for the state to exhibit
sub-Poissonian statistics. We then study the effect of decoherence
on the single-photon subtracted squeezed states. We present results
for two different models of decoherence, viz. amplitude decay model
and the phase diffusion model. In each case we give analytical
results for the time evolution of the state. We discuss the loss of
nonclassicality as a result of decoherence. We show through the
study of their phase-space properties how these states decay to
vacuum due to the decay of photons. We show that phase damping leads
to very slow decoherence than the photon-number decay.
\end{abstract}
\pacs{03.65.Yz,42.50.Dv}
\maketitle
\begin{figure*}
\caption{\label{vacuum}(a) Gaussian Wigner function $W_{\rm sq}$ of
the squeezed vacuum state $|\psi\rangle$ for $\theta=0$ and
$r=0.31$. (b) The Gaussian Wigner function of a vacuum state.}
\end{figure*}
\section{Introduction}
Quantum states can be well described in terms of Wigner functions.
The states with Gaussian Wigner function have been of particular
interests in context of quantum information processing. Entanglement
in such states in terms of quadratures are also well studied. On the
other hand, the quantum states with non-Gaussian Wigner function are
also quite important. For example, a single-photon state, which
finds many applications in quantum information processing,  shows
non-Gaussian behavior in phase-space. In a recent experiment, a
non-Gaussian state has been produced by homodyne detection technique
from a single-mode squeezed state of light \cite{wenger,kim}.
For certain non-Gaussian states, Wigner functions can take negative
values. Such negativity refers to nonclassicality of these states.
These states are useful in entanglement distillation
\cite{eisert,cirac}, loophole-free tests of Bell's inequality
\cite{bell}, and quantum computing \cite{sanders}. A specific class
of such nonclassical states has been shown to be similar to the
Schrodinger kitten state, in the sense that their Wigner functions
show negativity at the origin of phase space \cite{science,polzik}.
It is well known that the Schrodinger cat states \cite{knight},
which are quantum superpositions of coherent states, are
non-classical in nature and are very important to study the interface
of quantum and classical worlds. Superposition of coherent states
with low amplitudes creates Schrodinger kitten states. Most of the
experiments to prepare the Schrodinger cat states have been
performed in cavities or bound systems. Thus they are not much
useful in quantum information networks though they have the
non-Gaussian nature which is required in certain quantum
communication protocols. In \cite{science,polzik}, it has been shown
how to prepare an Schrodinger kitten state in an optical system, by
subtracting a single photon from a squeezed vacuum state. This
optical kitten state would overcome the limitations of bound
systems. Repeated photon-subtractions can lead to conditional
generation of arbitrary single-mode state \cite{cerf}. We note that
similar non-Gaussian states could be prepared by adding a single
photon to a squeezed vacuum state (see \cite{tara,bellini} for
details of photon-added coherent states, which are also non-Gaussian
states). These states are equivalent to single-photon subtracted
squeezed vacuum state and exhibit similar behavior in phase space.
It is worth to mention that non-Gaussian two-mode entangled states
can be prepared by subtracting a photon from a two-mode squeezed
state \cite{sasaki,paris}.

\begin{figure*}
\begin{center}
\caption{\label{sq_figs}Plots of Wigner functions of single-photon
subtracted squeezed states for (a) $r=0.31$ and (b) $r=0.8$ with
$\theta=0$.}
\end{center}
\end{figure*}
\begin{figure}
\caption{\label{cond}Variation of $C$ in phase space for $r=0.31$
and $\theta=0$.}
\end{figure}
\begin{figure}
\begin{center}
\caption{\label{ellipse}Contour plot for $C=$ constant in phase space for
(a) $r=0.31$ and (b) $r=0.8$.}
\end{center}
\end{figure}

In this paper, we focus our study on the nonclassical properties and
decoherence of single-photon subtracted squeezed vacuum states which
are optically produced single-mode non-Gaussian states. The
structure of the paper is as follows. In Sec. II, we introduce the
photon-subtracted squeezed states and discuss its nonclassical
properties in terms of the sub-Poissonian statistics and the
negativity of its Wigner function. We derive a compact expression of
the Wigner function and find the region in phase space where it
becomes negative. We show that there is an upper bound of the
squeezing parameter for this state to exhibit sub-Poissonian
statistics. In Sec. III, we study the effects of two different model
of decoherence: photon-number decay and phase damping. In both
cases, we derive analytical expressions for the time-evolution of
the state and its Wigner function. We discuss the loss of
nonclassicality due to decoherence. We show through the study of
evolution of the Wigner function how the state decays to vacuum as a
result of photon-number decay. We further show that phase damping
leads to much slower decoherence than the photon-number decay.

\begin{figure}
\begin{center}
\scalebox{0.6}{\includegraphics{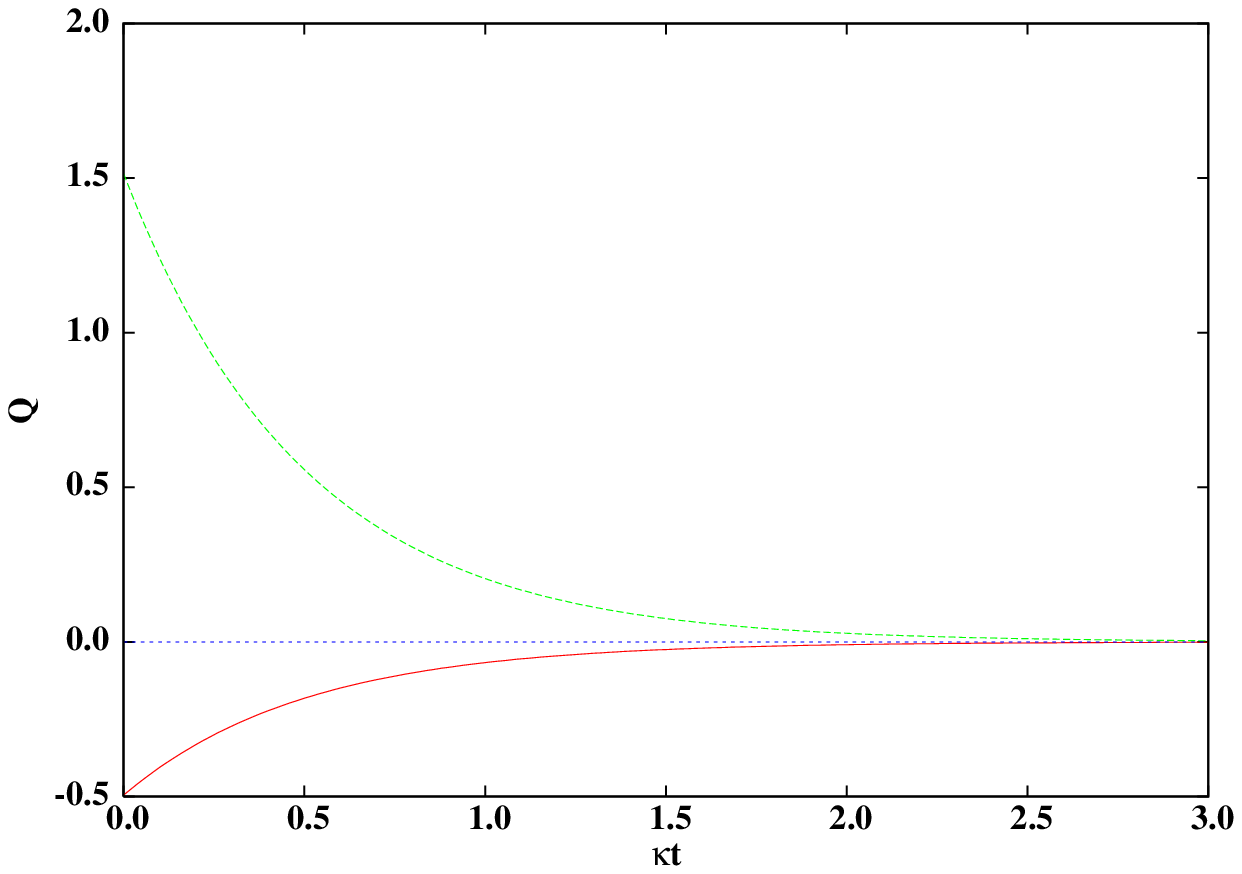}} 
\end{center}
\caption{\label{Qdecay}Variation of the $Q$-parameter
with time in presence of decoherence due to decay of
photon for squeezing parameters $r=0.31$ (red line) and $r=0.8$
(green line).}
\end{figure}

\begin{figure}
\caption{\label{negative_time}Variation of $P/(a^2-4|c|^2)^{5/2}$ in
phase space for $r=0.31$, $\theta=0$, and $\kappa t=0.1$.}
\end{figure}

\section{Photon-subtracted squeezed states}
An unnormalized single-mode squeezed vacuum state is given by
\begin{equation}
|\psi\rangle\equiv
S(\zeta)|0\rangle\;,\;S(\zeta)=\exp\left[\frac{\zeta}{2} a^{\dag
2}-\frac{\zeta^*}{2} a^2\right]\;,
\end{equation}
where $S(\zeta)$ is the squeezing operator, $\zeta=re^{i\theta}$ is
the complex squeezing parameter, and $a$ is the annihilation
operator. The Wigner function of this state is Gaussian and positive
in phase space. When $p~(>0)$ number of photons are subtracted from
such states, the state can be written as
\begin{equation}
|\psi\rangle_p \equiv a^p S(\zeta)|0\rangle \equiv
a^p\exp\left[\frac{\xi}{2}a^{\dag 2}\right]|0\rangle\;,
\end{equation}
where $\xi=\tanh(\zeta)$. For odd $p=2m+1$, the normalized form of
this state can be written as
\begin{equation}
\label{odd}|\psi\rangle_p =\frac{1}{N_o}\sum_{s=0}^\infty
\frac{(\xi/2)^{s+m+1}}{(s+m+1)!}\frac{(2s+2m+2)!}{\sqrt{2s+1}}|2s+1\rangle\;,
\end{equation}
while for even $p=2m$, the state becomes
\begin{equation}
\label{even}|\psi\rangle_p=\frac{1}{N_e}\sum_{s=0}^\infty
\frac{(\xi/2)^{s+m}}{(s+m)!}\frac{(2s+2m)!}{\sqrt{2s}}|2s\rangle\;,
\end{equation}
where $N_o$ and $N_e$ are the normalization constants. In this
paper, we focus on the case when a single photon is subtracted from
the squeezed vacuum, i.e., for $m=0$ in (\ref{odd}).
\begin{figure}
\begin{center}
\caption{\label{ellipse1}Plot of $16C'/(1-e^{-2\kappa t})$ in phase
space for (a) $r=0.31$ and (b) $r=0.8$ at times (i) $\kappa t=0.05$,
(ii) $\kappa t=0.1$, (iii) $\kappa t=0.3$, and (iv) $\kappa t=0.5$.}
\end{center}
\end{figure}

\begin{figure*}
\begin{center}
\caption{\label{small_sq}Wigner function of single-photon
subtracted squeezed states for $\theta=0$ and $r=0.31$ at (a)
$\kappa t=0.05$, (b) $\kappa t=0.1$, (c) $\kappa t=0.3$, and (d)
$\kappa t=0.5$.}
\end{center}
\end{figure*}

\begin{figure*}
\begin{center}
\caption{\label{large_sq}Wigner function of single-photon
squeezed states for $\theta=0$ and $r=0.8$ at (a) $\kappa t=0.1$,
(b) $\kappa t=0.3$, (c) $\kappa t=0.5$, and (d) $\kappa t=0.7$.}
\end{center}
\end{figure*}
\subsection{Negativity of the Wigner function}
The Wigner function of the squeezed vacuum state $|\psi\rangle$ is
given by
\begin{equation}
\label{w_sq}W_{\rm
sq}(\alpha,\alpha^*)=\frac{2}{\pi}\exp(-2|\tilde{\alpha}|^2)\;,
\end{equation}
where $\tilde{\alpha}=\alpha\cosh(r)-\alpha^*e^{i\theta}\sinh(r)$.
The function is Gaussian in phase space, as shown in Fig.
\ref{vacuum}. We now calculate the Wigner function of the state
$|\psi\rangle_p$. This state can be rewritten as
\begin{equation}
\label{rel}|\psi\rangle_p =a^p S(\zeta)|0\rangle=S(\zeta)S^\dag
(\zeta)a^pS(\zeta)|0\rangle\;.
\end{equation}
Using the relation
\begin{equation}
S^\dag(\zeta)a^pS(\zeta)=[\cosh(r)a+e^{i\theta}\sinh(r)a^{\dag}]^p\;,
\end{equation}
we get the following from (\ref{rel})
\begin{equation}
\label{state}|\psi\rangle_1\equiv S(\zeta)|1\rangle\;,
\end{equation}
for $p=1$. For a density matrix $\tilde{\rho}(a,a^\dag)$, we can
write the following:
\begin{equation}
\label{relation}S(\zeta)\tilde{\rho}(a,a^\dag)S^\dag
(\zeta)=\tilde{\rho}[S(\zeta)aS^\dag (\zeta),S(\zeta)a^\dag
S^\dag(\zeta)]\;.
\end{equation}
We write the Wigner function of $\tilde{\rho}(a,a^\dag)$ as
$W_{\tilde{\rho}}(\alpha,\alpha^*)$. Using the identities
\begin{eqnarray}
S(\zeta)aS^\dag (\zeta) &=&a\cosh(r)-a^\dag e^{i\theta}\sinh(r)\;,\\
S(\zeta)a^\dag S^\dag(\zeta) &=& a^\dag
\cosh(r)-ae^{-i\theta}\sinh(r)\;,
\end{eqnarray}
in (\ref{relation}), we can thus write the Wigner function of the
density matrix $\rho(a,a^\dag)=S(\zeta)\tilde{\rho}(a,a^\dag)S^\dag
(\zeta)$ as
\begin{equation}
\label{formula}W_{\rho}(\alpha,\alpha^*)=W_{\tilde{\rho}}(\tilde{\alpha},\tilde{\alpha}^*)\;,
\end{equation}
where we have used the linearity property of the Wigner function.
For the state (\ref{state}), $\tilde{\rho}=|1\rangle\langle 1|$ and
its Wigner function is given by
\begin{equation}
\label{wsq}W_{\tilde{\rho}}(\alpha,\alpha^*)=\frac{2}{\pi}(4|\alpha|^2-1)e^{-2|\alpha|^2}\;.
\end{equation}
Thus, the Wigner function of the single-photon subtracted squeezed
vacuum state becomes
\begin{equation}
\label{wig1}W_\rho(\alpha,\alpha^*)=\frac{2}{\pi}(4|\tilde{\alpha}|^2-1)e^{-2|\tilde{\alpha}|^2}\;,
\end{equation}
where we have used (\ref{formula}) and (\ref{wsq}). Clearly, the
Wigner function (\ref{wig1}) is non-Gaussian in phase-space. We show
the plot of this Wigner function in the phase-space in Figs.
\ref{sq_figs} for different squeezing parameters. As an evidence of
non-classicality of the state, squeezing in one of the quadratures
is clear in the plots. Also there is some negative region of the
Wigner function in the phase-space which is another evidence of the
non-classicality of the state. The function becomes negative in
phase space, when
\begin{equation}
\label{neg_cond}|\tilde{\alpha}|^2<\frac{1}{4}\;.
\end{equation}
We show in Fig. \ref{cond} the variation of
$C=|\tilde{\alpha}|^2-\frac{1}{4}$ in phase space. The negative
region of $C$ corresponds to the negativity of the Wigner function.
Note that $C=$ constant corresponds to ellipse in phase space, as
shown in Fig. \ref{ellipse}.

Note that the photon-subtracted squeezed states are similar to
Schrodinger kitten states \cite{science,polzik} because Wigner functions
exhibit the same characteristics in phase-space especially for the
large values of the squeezing parameters. Moreover, in both cases,
Wigner function becomes negative in the center of phase space.

\subsection{Sub-Poissonian nature of the photon-subtracted state}
The nonclassicality of the state $|\psi\rangle_1$ can also be
analyzed by studying its sub-Poissonian character in terms of the
Mandel's $Q$-parameter \cite{mandel} which is defined by
\begin{equation}
\label{Qpara}Q=\frac{\langle a^{\dag 2}a^2\rangle-\langle a^\dag
a\rangle^2}{\langle a^\dag a\rangle}\;.
\end{equation}
The negativity of the $Q$-parameter refers to sub-Poissonian
statistics of the state. However in \cite{tara_Q}, it has been shown
that a state can be nonclassical even if $Q$ is positive. A similar
situation occurs in the present case. For the state (\ref{odd}), we
find
\begin{eqnarray}
\langle a^{\dag
2}a^2\rangle &=&\sum_{n=0}^\infty n(n-1)\rho_{n,n}=\frac{3|\xi|^4(3+2|\xi|^2)}{N_o^2(1-|\xi|^2)^{7/2}}\;,\nonumber\\
\label{avg1}\langle a^\dag a\rangle &=&\sum_{n=0}^\infty
n\rho_{n,n}=\frac{|\xi|^2(1+2|\xi|^2)}{N_o^2(1-|\xi|^2)^{5/2}}\;,
\end{eqnarray}
where the normalization constant is given by
\begin{equation}
N_o^2=\frac{|\xi|^2}{(1-|\xi|^2)^{3/2}}\;.
\end{equation}
From (\ref{avg1}), we find that $Q$ becomes negative for
$|\xi|\lesssim 0.43$, which is satisfied for $r\lesssim 0.46$. We
emphasize that the Wigner function has negative region for all
values of $r$, and thus the photon-subtracted squeezed state is
nonclassical for all $r$, though it does not exhibit sub-Poissonian
photon statistics above certain squeezing threshold.

\section{Models of decoherence}
We next consider how this state evolves under decoherence. The
decoherence of the single-mode state (\ref{odd}) can be due to decay
of photons to the reservoir or due to phase damping.

\subsection{Amplitude decay model}
When the photons decay to reservoir, the corresponding Markovian
dynamics of the state is well described by the following equation:
\begin{equation}
\frac{d}{dt}\rho=-\kappa(a^\dag a\rho-2a\rho a^\dag+\rho a^\dag a)\;,
\end{equation}
where $\kappa$ is the rate of decay. The solution of this equation
can be written as
\begin{equation}
\label{sol}\rho(t)=\sum_{n,n'}\rho_{n,n'}(t)|n\rangle\langle n'|\;,
\end{equation}
where the density matrix element $\rho_{n,n'}(t)$ can be found by
using the Laplace transformation and the iteration methods
\cite{gsa_po}. To see this, let us start with the time-dependent
equation for $\rho_{n,n'}$:
\begin{equation}
\dot{\rho}_{n,n'}=-\kappa(n+n')\rho_{n,n'}+2\kappa\sqrt{(n+1)(n'+1)}\rho_{n+1,n'+1}\;.
\end{equation}
Using the new subscripts $q=n-n'$ and $p=(n+n')/2$, the above
equation transforms into
\begin{equation}
\label{pq}\dot{\rho}_{p,q}=-2\kappa p\rho_{p,q}+2\kappa
\sqrt{(p+1)^2-(q/2)^2}\rho_{p+1,q}\;.
\end{equation}
Taking Laplace transformation of (\ref{pq}) and using the original
subscript $n$ and $n'$, we can write the time-dependent solution for
the density matrix elements as
\begin{eqnarray}
\rho_{n,n'}(t)&=&e^{-\kappa
t(n+n')}\sum_{r=0}^\infty\sqrt{\left(^{n+r}C_r\right)\left(^{n'+r}C_r\right)}\nonumber\\
\label{rhonn}&&\times (1-e^{-2\kappa t})^r\rho_{n+r,n'+r}(t=0)\;,
\end{eqnarray}
where for the single-photon subtracted squeezed vacuum
\begin{eqnarray}
\rho_{n+r,n'+r}(t=0)&=&\frac{1}{N_o^2}\frac{(\xi/2)^{(n+r+1)/2}(\xi^*/2)^{(n'+r+1)/2}}{(\frac{n+r+1}{2})!(\frac{n'+r+1}{2})!}\nonumber\\
&&\label{init}\times\frac{(n+r+1)!(n'+r+1)!}{\sqrt{(n+r)!(n'+r)!}}\;.
\end{eqnarray}

We next calculate the parameter $Q$ [Eq. (\ref{Qpara})] for the
state (\ref{sol}). We have found that
\begin{eqnarray}
\langle a^{\dag 2}a^2\rangle &=&\sum_{n=0}^\infty
n(n-1)\rho_{n,n}(t)\;,\nonumber\\
\label{avg2}\langle a^\dag a\rangle &=&\sum_{n=0}^\infty n\rho_{n,n}(t)\;,
\end{eqnarray}
where $\rho_{n,n}(t)$ is given by (\ref{rhonn}) and (\ref{init}) for
$n=n'$ in case of state $|\psi\rangle_1$. Using Eqs. (\ref{avg2}), we
plot $Q$ with time in Fig. \ref{Qdecay}. It is easy to see that
at long times ($\kappa t\rightarrow \infty$), $Q$ vanishes. This is
because at this limit, $\rho_{n,n}(t)$ vanishes for all non-zero $n$
and $\rho_{0,0}(t\rightarrow \infty)=1$, i.e., the state decays to
vacuum. Thus the averages (\ref{avg2}) vanish and $Q$ also vanishes.

\subsubsection{Evolution of Wigner function}
The evolution of the Wigner function is governed by the following
equation:
\begin{equation}
\label{wignereq}\frac{\partial W}{\partial
t}=\kappa\left[\frac{\partial}{\partial
\alpha}\alpha+\frac{\partial}{\partial
\alpha^*}\alpha^*+\frac{\partial^2}{\partial
\alpha\partial\alpha^*}\right]W(\alpha,\alpha^*)\;.
\end{equation}
The solution can be written as
\begin{eqnarray}
\label{wignersol}W(\alpha,\alpha^*,t)&=&\frac{2}{\pi(1-e^{-2\kappa
t})}\int
d^2\alpha_0W(\alpha_0,\alpha_0^*,0)\nonumber\\
&&\exp\left\{-2\frac{|\alpha-\alpha_0e^{-\kappa
t}|^2}{(1-e^{-2\kappa t})}\right\}\;,
\end{eqnarray}
where $W(\alpha_0,\alpha_0^*,0)$ is the Wigner function of the
initial state. It is easy to verify this solution putting
(\ref{wignersol}) in (\ref{wignereq}). The time-evolution of the
Wigner function of the squeezed vacuum state $|\psi\rangle$ can be
easily calculated analytically  using the following integral
identity \cite{gsa_prdeq,puri}:
\begin{eqnarray}
\label{ident}&&\int
d^2\alpha\exp[-|\alpha|^2]\exp\left(-\frac{\mu}{\tau}\alpha^2-\frac{\nu}{\tau}\alpha^{*2}-\frac{z^*\alpha}{\sqrt{\tau}}+\frac{z\alpha^*}{\sqrt{\tau}}\right)\nonumber\\
&&=\frac{\pi\tau}{\sqrt{\tau^2-4\mu\nu}}\exp\left(-\frac{\mu
z^2+\nu z^{*2}+\tau|z|^2}{\tau^2-4\mu\nu}\right)\;.
\end{eqnarray}
Using Eq. (\ref{wsq}) and the above identity in Eq.
(\ref{wignersol}), we get the following:
\begin{eqnarray}
W(\alpha,\alpha^*,t)&=&\frac{4}{\pi(1-e^{-2\kappa t})}
\frac{\exp\left[-\frac{2|\alpha|^2}{1-e^{-2\kappa t}}\right]}{\sqrt{a^2-4|c|^2}}\nonumber\\
&&\times\exp\left[\frac{b^2c^*+b^{*2}c+a|b|^2}{a^2-4|c|^2}\right]\;,
\end{eqnarray}
where
\begin{eqnarray}
a&=&2\cosh(r)+2\frac{e^{-2\kappa t}}{1-e^{-2\kappa t}}\;,\nonumber\\
b&=&\frac{2\alpha^*e^{-\kappa t}}{1-e^{-2\kappa
t}}\;,\;c=e^{-i\theta}\sinh(r)\;.
\end{eqnarray}
Clearly the Wigner function of the squeezed vacuum state is Gaussian
at all times.

We now calculate the time-dependence of the Wigner function of the
state $|\psi\rangle_1$. The initial Wigner function, as given by
(\ref{wig1}), can be rewritten as
\begin{equation}
\label{alter_W}W(\alpha,\alpha^*,0)=\frac{2}{\pi}D\left.\left[e^{-\lambda|\tilde{\alpha}|^2}\right]\right|_{\lambda=2}\;,\;D=-4\frac{d}{d\lambda}-1\;.
\end{equation}
Using (\ref{alter_W}) and (\ref{wignersol}), we can find the
following expression for the Wigner function:
\begin{eqnarray}
W(\alpha,\alpha^*,t)&=&\left(\frac{2}{\pi}\right)^2\frac{1}{1-e^{-2\kappa
t}}D\int
d^2\alpha_0\exp[-\lambda|\tilde{\alpha}_0|^2]\nonumber\\
&&\times \left.\exp\left[-2\frac{|\alpha-\alpha_0e^{-\kappa
t}|^2}{1-e^{-2\kappa t}}\right]\right|_{\lambda=2}\;,
\end{eqnarray}
where
$\tilde{\alpha}_0=\alpha_0\cosh(r)-\alpha_0^*e^{i\theta}\sinh(r)$.
Simplifying the above expression using Eq. (\ref{ident}), we get
\begin{eqnarray}
W(\alpha,\alpha^*,t)&=&\frac{32P}{\pi(1-e^{-2\kappa
t})}\frac{\exp\left[-\frac{2|\alpha|^2}{1-e^{-2\kappa
t}}\right]}{(a^2-4|c|^2)^{5/2}}\nonumber\\
&&\label{timeW}\times\exp\left[\frac{b^2c^*+b^{*2}c+a|b|^2}{a^2-4|c|^2}\right]\;,
\end{eqnarray}
where
\begin{eqnarray}
P&=&(1-x^2)\{\sinh(2r)(e^{i\theta}b^2+e^{-i\theta}b^{*2})\nonumber\\
&&+2[(x+1)^2+4x\sinh^2(r)]\}\nonumber\\
\label{P}&&+2\{(1+x^2)\cosh(2r)+2x\}|b|^2
\end{eqnarray}
and
\begin{equation}
x=\frac{e^{-2\kappa t}}{1-e^{-2\kappa t}}\;.
\end{equation}
Clearly, the Wigner function is non-Gaussian due to the presence of
the polynomial $P$. This becomes negative when the polynomial $P$
becomes negative. In Fig. \ref{negative_time}, we have plotted
$C'(\alpha,\alpha^*,t)=P/(a^2-4|c|^2)^{5/2}$ in phase space to show
the negative region for Wigner function.

Note that at the center of the phase space ($\alpha= \alpha^*=0$),
the Wigner function is maximally negative. At the center,
\begin{equation}
C'(0,0,t)=\frac{2(1-x^2)\{(x+1)^2+4x\sinh^2(r)\}}{(a^2-4|c|^2)^{5/2}}\;,
\end{equation}
which becomes negative when $(1-x^2)$ becomes negative. This leads
to the following condition:
\begin{equation}
\kappa t<\kappa t_0=\frac{1}{2}\ln(2)\;,
\end{equation}
which is independent of the squeezing parameter $r$. Thus the Wigner
function has certain negative region for the time
$t<t_0=\ln(2)/2\kappa$. However the negative value does depend on
the squeezing parameter.

We find from Eq. (\ref{P}) that at $t=t_0$ (i.e, when $x^2=1$), $P$
becomes a circle in phase space. Thus beyond $t>t_0$, the ellipse
$16C'/(1-e^{-2\kappa t})=$ constant interchanges its minor and major
axes. We show this behavior in Figs. \ref{ellipse1} for different
values of $r$. Note that at times much larger than decoherence
time-scale $1/\kappa$ (i.e., for $\kappa t\rightarrow \infty$),
$P\rightarrow 2$ and thus becomes constant throughout the phase
space.

Using Eq. (\ref{timeW}), we show the variation of Wigner function at
different time-scales in Figs. \ref{small_sq}. It is easy to see how
the negative region of the Wigner function gradually diminishes. At
long times $\kappa t\rightarrow \infty$, the Wigner function becomes
\begin{equation}
W(\alpha,\alpha^*,\infty)= \frac{2}{\pi}e^{-2|\alpha|^2}\;,
\end{equation}
which corresponds to vacuum state. We have shown this in Fig.
\ref{vacuum}(b). This can also be understood from Eq. (\ref{rhonn}).
For $\kappa t\rightarrow \infty$, $\rho_{0,0}$ approaches unity,
whereas all other density matrix elements vanish. This means that at
long times, the state decays to vacuum, as we have discussed
earlier.

We next study the time-evolution of the Wigner function for the case
of large squeezing, i.e., large values of $\zeta$. In this case the
single photon subtracted squeezed state becomes similar to a
Schrodinger cat state. For large times, such an optical cat state
decays to vacuum. Thus the Wigner function becomes Gaussian, as
discussed above. We show this evolution for large squeezing in Figs.
\ref{large_sq}.

\subsection{Effect of phase damping}
We now study the effect of phase-damping on the state
$|\psi\rangle_1$. Such damping can be described by the following
master equation:
\begin{equation}
\dot{\rho}=-\kappa_p(A^{\dag}A\rho-2A\rho A^\dag+\rho A^\dag A)\;,
\end{equation}
where $A=a^\dag a$ is the number operator and $\kappa_p$ is the
corresponding rate of decoherence. The solution of this equation can
be easily found as (\ref{sol}) where
\begin{equation}
\label{sol_phase}\rho_{n,n'}(t)=\exp[-(n-n')^2\kappa_p
t]\rho_{n,n'}(0)\;.
\end{equation}
It is easy to see that only the diagonal elements $\rho_{n,n}$ do
not decay due to dephasing. Thus at long times, we can write
\begin{equation}
\rho(t\rightarrow \infty)=\sum_{n=0}^\infty
\rho_{n,n}(0)|n\rangle\langle n|\;,
\end{equation}
which refers to a mixed state.

Using Eqs. (\ref{avg2}) and (\ref{sol_phase}), we next calculate the
parameter $Q$. We find that the averages
$\langle a^{\dag 2}a^2\rangle$ and $\langle a^\dag a\rangle$ do not
depend upon time, because in case of phase damping
$\rho_{n,n}(t)=\rho_{n,n}(0)$. Thus $Q$ remains the same for all
times.
\begin{figure*}
\begin{center}
\caption{\label{wig_phase_fig}Wigner function in phase
space at long times in presence of phase damping for (a) $r=0.31$
and (b) $r=0.8$.}
\end{center}
\end{figure*}
However, the corresponding Wigner function has certain
time-dependence. We find that at long times, the Wigner function
becomes
\begin{equation}
\label{wigner_phase}W(\alpha,\alpha^*,\infty)=\sum_{n=0}^\infty
\rho_{n,n}(0)W_{|n\rangle\langle n|}(\alpha,\alpha^*)\;,
\end{equation}
where $W_{|n\rangle\langle n|}(\alpha,\alpha^*)$ is the Wigner
function of a Fock state $|n\rangle$ as given by
\begin{equation}
W_{|n\rangle\langle
n|}(\alpha,\alpha^*)=(-1)^n\frac{2}{\pi}e^{-2|\alpha|^2}L_n(4|\alpha|^2)\;.
\end{equation}
The function (\ref{wigner_phase}) refers to a highly nonclassical
state. It is interesting to note that all the Fock states have
independent contributions to the Wigner function at long times,
weighted by their initial population $\rho_{n,n}(0)$. On the other
hand, in case of decoherence due to photon-number decay, only the
vacuum state survives. In Fig. \ref{wig_phase_fig}, we plot the
Wigner function (\ref{wigner_phase}) in phase space for different
squeezing. Note that the Wigner function has negative region at long
times representing nonclassicality for all $r$, even if the state
does not exhibit sub-Poissonian statistics for $r\gtrsim 0.46$
(because $Q$ is positive). In fact, if $Q$ is positive, it does not
mean that the the state is classical. In such cases, we have to use
other parameters to test the nonclassicality. Several parameters
have been introduced in this context \cite{tara_Q,nonclass}. We can use
hierarchy of these parameters which have been shown to be especially
useful in context of cat states. Here we illustrate the utility of one such
parameter, e.g., the $A_3$ parameter as defined by \cite{tara_Q}
\begin{equation}
\label{a3}A_3=\frac{{\rm det}[m^{(3)}]}{{\rm det}[\mu^{(3)}]-{\rm
det}[m^{(3)}]}\;,
\end{equation}
where
\begin{equation}
m^{(3)}=\left(\begin{array}{ccc} 1&m_1&m_2\\
m_1&m_2&m_3\\
m_2&m_3&m_4
\end{array}\right)\;,\;\mu^{(3)}=\left(\begin{array}{ccc} 1&\mu_1&\mu_2\\
\mu_1&\mu_2&\mu_3\\
\mu_2&\mu_3&\mu_4
\end{array}\right)\;,
\end{equation}
$m_s=\langle a^{\dag s}a^s\rangle$, $\mu_s=\langle (a^\dag
a)^s\rangle$ and det indicates determinant of the matrix. The state exhibits phase-insensitive nonclassical
properties if $A_3$ lies between 0 and -1 \cite{tara_Q}. For the
state $|\psi\rangle_1$ we have found that $A_3$ remains negative for
$|\xi|\lesssim 0.6$ which corresponds to $r\lesssim 0.7$. Clearly
$A_3$ is a stronger measure of nonclassicality than $Q$ because it
leads to a larger upper bound of $r$ to exhibit nonclassicality.
Further, comparing the Wigner functions in Figs. \ref{wig_phase_fig} with those at $t=0$ [see
Figs. \ref{sq_figs}], we find that the Wigner function varies very
slowly with time for small squeezing. But for large squeezing, the
variation is faster. Although we can conclude that phase damping
leads to much slower decoherence than amplitude damping.

\section{Conclusions}
In conclusion, we have studied how a class of non-Gaussian states
evolves in presence of decoherence. We have considered a
single-photon subtracted squeezed vacuum state, the Wigner function
of which is similar to that of a Schrodinger kitten state. We have
found an upper bound for squeezing parameter for which this state
exhibits sub-Poissonian photon statistics. However, the state
remains nonclassical for all values of the squeezing parameter
because the Wigner function becomes negative around central region
in phase space. Next, we have studied how the state evolves in
presence of two different kinds of decoherence, viz., amplitude
decay and phase damping. We have found analytical expressions for
the time-evolution of the state and the Wigner function in both
cases. In case of amplitude decay, the Wigner function loses its
non-Gaussian nature and becomes Gaussian at long times,
corresponding to vacuum. On the other hand, phase damping leads to
much slower decoherence than amplitude damping. The state remains
nonclassical at long times.

\begin{acknowledgments}
A.B. gratefully acknowledges the partial support from the Women in
Science and Engineering program in University of Southern
California, Los Angeles, USA. G.S.A. kindly acknowledges support
from NSF grant no. CCF0524673.
\end{acknowledgments}


\begin{thebibliography}{999}

\bibitem{wenger}
J. Wenger, R. Tualle-Brouri, and P. Grangier, Phys. Rev. Lett. {\bf
92}, 153601 (2004).

\bibitem{kim}
M. S. Kim, E. Park, P. L. Knight, and H. Jeong, Phys. Rev. A {\bf
71}, 043805 (2005).

\bibitem{eisert}
J. Eisert, S. Scheel, and M. B. Plenio, Phys. Rev. Lett. {\bf 89},
137903 (2002); D. E. Browne, J. Eisert, S. Scheel, and M. B. Plenio,
Phys. Rev. A {\bf 67}, 062320 (2003).

\bibitem{cirac}
G. Giedke and J. I. Cirac, Phys. Rev. A {\bf 66}, 032316 (2002).

\bibitem{bell}
H. Nha and H. J. Carmichael, Phys. Rev. Lett. {\bf 93}, 020401
(2004); R. Garcia-Patron, J. Fiurasek, N. J. Cerf, J. Wenger, R.
Tualle-Brouri, and P. Grangier, Phys. Rev. Lett. {\bf 93}, 130409
(2004).

\bibitem{sanders}
S. D. Bartlett and B. C. Sanders, Phys. Rev. A {\bf 65}, 042304
(2002).

\bibitem{science}
A. Ourjoumtsev, R. Tualle-Brouri, J. Laurat, and P. Grangier,
Science {\bf 312}, 83 (2006).

\bibitem{polzik}
J. S. Neergaard-Nielsen, B. Melholt-Nielsen, C. Hettich, K. M\o
lmer, and E. S. Polzik, Phys. Rev. Lett. {\bf 97}, 083604 (2006).

\bibitem{knight}
V. Buzek and P. L. Knight, {\it Progress in Optics\/}. Ed. E. Wolf
(North Holland, Amsterdam, 1195), vol. XXXIV, 1.

\bibitem{cerf}
J. Fiurasek, R. Garcia-Patron, and N. J. Cerf, Phys. Rev. A {\bf
72}, 033822 (2005).

\bibitem{tara}
G. S. Agarwal and K. Tara, Phys. Rev. A {\bf 43}, 492 (1991).

\bibitem{bellini}
A. Zavatta, S. Viciani, and M. Bellini, Science {\bf 306}, 660
(2004); Phys Rev. A {\bf 72}, 023820 (2005).

\bibitem{sasaki}
M. Sasaki and S. Suzuki, Phys. Rev. A {\bf 73}, 043807 (2006).

\bibitem{paris}
C. Invernizzi, S. Olivares, M. G. A. Paris, and K. Banaszek, Phys.
Rev. A {\bf 72}, 042105 (2005); S. Olivares and M. G. A. Paris, {\it
ibid.\/} {\bf 70}, 031112 (2004); S. Olivares, M. G. A. Paris, and
R. Bonifacio, {\it ibid.\/} {\bf 67}, 032314 (2003).

\bibitem{mandel}
L. Mandel, Opt. Lett. {\bf 4}, 205 (1979).

\bibitem{tara_Q}
G. S. Agarwal and K. Tara, Phys. Rev. A {\bf 46}, 485 (1992).

\bibitem{gsa_po}
G. S. Agarwal, {\it Progress in Optics\/}, Ed. E. Wolf (North
Holland, Amsterdam, 1973), vol. XI, 1.

\bibitem{gsa_prdeq}
G. S. Agarwal and E. Wolf, Phys. Rev. D {\bf 2}, 2187 (1970), Eq.
(6.15); V. Bargmann, Commun. Pure Appl. Math. {\bf 14}, 187 (1961).

\bibitem{puri}
R. R. Puri, {\it Mathematical Methods of Quantum Optics}
(Springer-Verlag, Berlin, 2001), Appendix A.

\bibitem{nonclass}
G. S. Agarwal, Opt. Commun. {\bf 93}, 109 (1993); W. Vogel, Phys. Rev. Lett. {\bf 84}, 1849 (2000); Th. Richter and W.
Vogel, {\it ibid.\/} {\bf 89}, 283601 (2002).

\end{thebibliography}
\end{document}